\documentclass[aps,prl,reprint,groupedaddress,showpacs]{revtex4-1}

\usepackage{graphicx}
\usepackage{dcolumn}
\usepackage{bm}
\usepackage{color} 
\usepackage{hyperref}
\usepackage{amsmath}

\newcommand{\xold}[1]{\textcolor{black}{#1}}
\newcommand{\xmg}[1]{\textcolor{black}{#1}}

\begin{document}


\title{Rheological Chaos of Frictional Grains}

\author{Matthias Grob, Annette Zippelius and Claus Heussinger}
\affiliation{Institute of Theoretical Physics, Georg-August University of G\"ottingen, 37073 G\"ottingen, Germany}


\date{\today}

\begin{abstract}
  A two-dimensional dense fluid of frictional grains is shown to
  exhibit time-chaotic, spatially heterogeneous flow in a range of
  stress values, $\sigma$, chosen in the unstable region of s-shaped
  flow curves. Stress controlled simulations reveal a phase diagram
  with reentrant stationary flow for small and large stress
  $\sigma$. In between no steady flow state can be reached, instead
  the system either jams or displays time dependent heterogeneous
  strain rates $\dot\gamma({\bf r},t)$. The results of
  simulations are in agreement with the stability analysis of a simple
  hydrodynamic model, coupling stress and microstructure which we
  tentatively associate with the frictional contact network.
 \end{abstract}


\maketitle


Discontinuous shear thickening is a ubiquitous phenomenon, observed in many dense
suspensions~\cite{JaegerReview,Bonn}. Simulations
for non-Brownian particle suspensions~
\cite{fernandez2013,heussinger2013,Mari2013} as well as for frictional
granular media~\cite{otsuki2011critical,Grob2014} have highlighted the
particular role played by frictional particle
interactions. Discontinuous shear thickening implies a region of shear
stress which (at finite Reynolds number, Ref.~\cite{Mari2015}) is not
accessible to a homogeneous system in stationary state. What happens,
if the system is forced into this regime by prescribing the stress at
the boundary in the unstable region? One possibility is vorticity
banding~\cite{Olmsted2008}, corresponding to bands with different
stress values at the same shear rate.  However there is no clear
evidence for persistent vorticity banding in experiment so far. Furthermore
objections have been raised as to the possibility of vorticity banding
as a stationary state: The pressure - in contrast to the shear stress
- has to be the same across the interface of the bands; otherwise
particle migration is expected to occur and thereby destabilise the
interface. If stationary states are not accessible to the system, we
expect to observe time-dependent, inhomogeneous states, either
oscillatory or chaotic~\cite{Cates2002}. 

In this paper we show that spatio-temporal chaos occurs in a two-dimensional system of frictional granular
particles, subject to an applied stress which is chosen in the
unstable region of the flow curve. We present results from simulations
and formulate a hydrodynamic model to derive a phase diagram and
identify the regions of parameter space, where time-chaotic,
inhomogeneous solutions are to be found.

We simulate a two-dimensional system of $N$ soft, frictional particles
in a square box of linear dimension $L$ as detailed
in~\cite{Grob2014}. The particles all have the same mass $m=1$, but
are polydisperse in size \xmg{with diameters $0.7$, $0.8$, $0.9$ and $1$ in equal amounts}. Normal and
tangential forces, $\bm{f}^{(n)}$ and $\bm{f}^{(t)}$, are modeled with
\xmg{linear spring-dashpots} of unit strength for both, elastic as
well as viscous contributions. Thereby units of time, length and mass
have been fixed\xmg{~\cite{Note3}. Flow curves for other visco-elastic parameters are presented in the} \xold{appendix}. Coulomb friction is implemented
with friction parameter $\mu=2$\xmg{, corresponding to the
  high friction limit. We expect the same qualitative findings as presented in this \xold{Rapid Communication} for all values of
  $\mu > 0$ and refer to a systematic study of the $\mu$-dependence in ref.~\cite{otsuki2011critical}.}
In the stress-controlled simulations, a boundary layer of particles is
frozen and the boundary at the top is moved with a force
$\sigma L\bm{\hat{e}}_x $, whereas the bottom plate remains at rest. \xmg{In shear direction we use periodic boundary conditions.}

{\bf Constitutive equation~-~}
Previous work~\cite{otsuki2011critical,Grob2014} has revealed
discontinuous shear thickening for a range of packing fractions 
close to the jamming transition.  The flow curves for frictional
granular particles are well represented by the following constitutive
equation
\begin{equation}
  \label{eq:fc1}
  \dot{\gamma} (\sigma) = a\sigma^{1/2} - b\sigma + c\sigma^2,
\end{equation}
where the term $-b\sigma$ is due to the frictional interactions \xmg{and dependence on the packing fraction, $\phi$, is implemented with $a=a(\phi)$~\cite{Note1,Grob2014}}. This gives rise to 
the phenomenology of the van-der-Waals theory for a first-order phase 
transition: jamming first occurs at the critical
point $\phi_c \cong 0.795$ 
which also marks the onset of hysteresis. A finite yield stress first
appears at $\phi_{\sigma} \cong 0.8003$, while the \xmg{generalized }viscosity\xmg{, $\eta = \sigma / \dot{\gamma}^2$,} diverges only at $\phi_{\eta} \cong 0.819$.
A similar, but not identical, sequence of characteristic \xmg{packing}
fractions has been proposed in
\cite{otsuki2011critical,ciamarra2011}. Most remarkably is the
existence (due to the $b$-dependent term) of a regime of shear stress
which is unstable $\frac{\partial\dot{\gamma}}{\partial\sigma}<0$,
corresponding to s-shaped flow curves. In strain-controlled conditions
such an s-shape leads to discontinuous shear thickening. In the
van-der-Waals theory it corresponds to the coexistence region.

\begin{figure}
 \includegraphics[angle=0, width=1.0\linewidth]{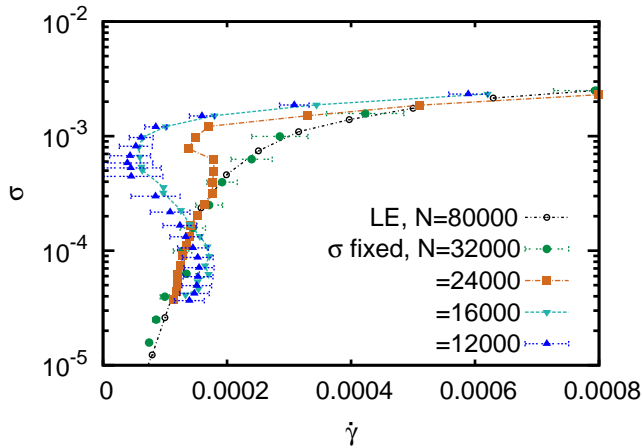}
 \caption{\label{fig:flowcurveS}Evolution of flow curves
   $\sigma(\dot\gamma)$ with system size $N$. \xmg{For small systems a
   pronounced s-shape is visible in transients before the system jams. In large systems no steady flow occurs; average over the time-dependent flow results in continuous shear thickening (stress controlled except for $N=80.000$).}}
\end{figure}

{\bf Simulations in the unstable regime~-~} These s-shaped flow curves
are indeed observed in the simulations (see Fig.~\ref{fig:flowcurveS})
however only as transients and only in rather small simulation cells \xmg{($N<24000$)}.
In larger systems the s-shape is slowly eroded and vanishes completely
above a certain system size. Instead a regime of continuous shear
thickening develops. Closer inspection of the simulations in this
regime reveals that no simple steady-state is reached. Rather, the
system displays time-chaotic and spatially inhomogeneous behavior. An
example is shown in Video~\ref{fig:chaotic_local_stress}, which is a
sequence of 4 snapshots of the local-stress field (movies are given in the supplementary material \cite{Note2}). The system does not settle
into a time-independent steady state on the timescale of the
simulations. Instead one observes time-dependent large scale
structures, e.g. shear bands which seem to propagate in the principal
stress direction, alternating with approximately homogeneous states
and random large scale structures. In Fig.~\ref{fig:chaotic_strain} we
show the corresponding time-dependent strain rate. One
clearly observes irregular time-dependence with intermittent
oscillatory periods.
\begin{video}
\includegraphics[angle=0, width=0.99\linewidth]{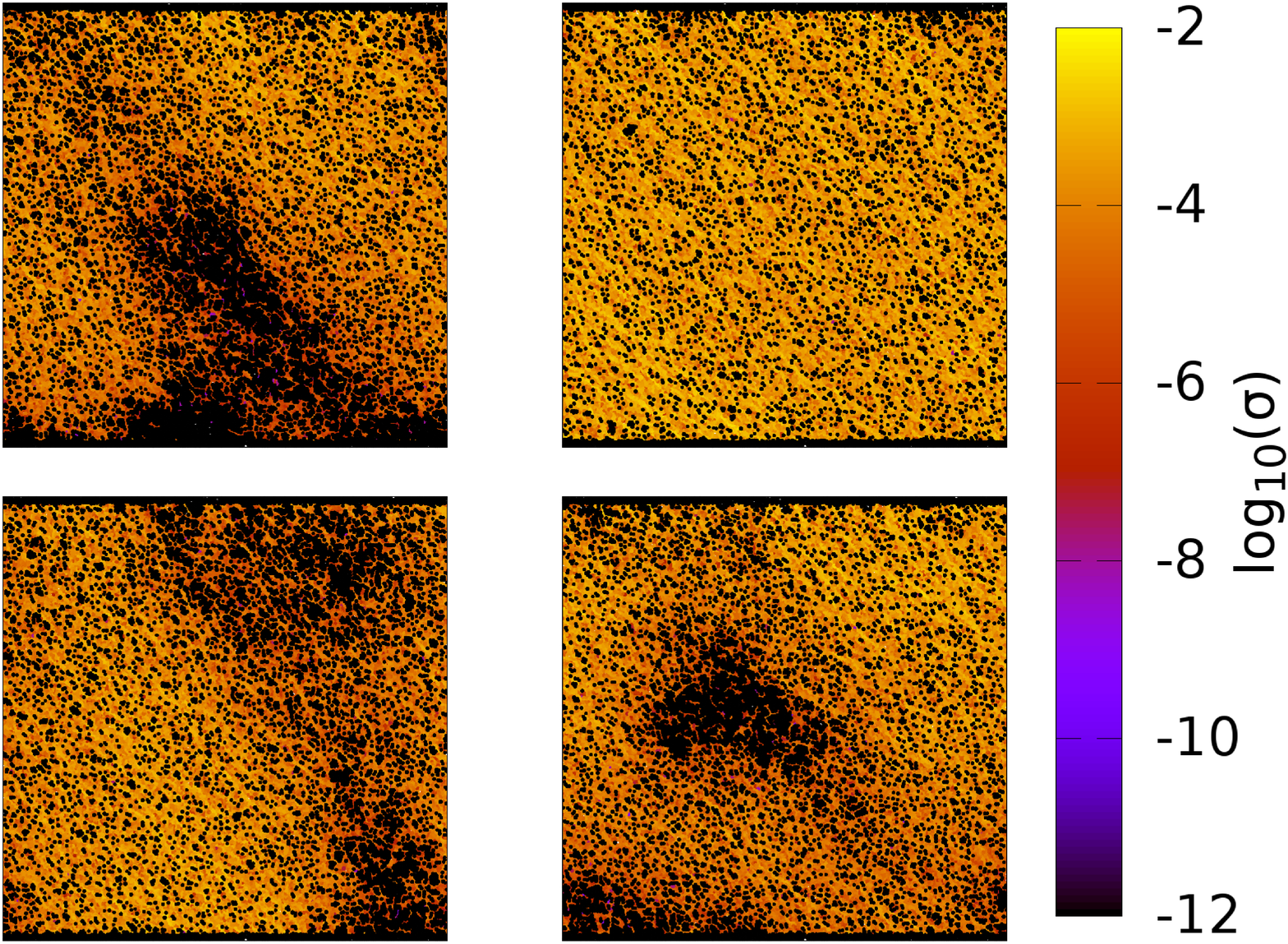}
\setfloatlink{To Supplemental Material}
 \caption{\label{fig:chaotic_local_stress} Four snapshots of the local
   stress, revealing large scale, time-dependent structures; $\phi = 0.8035$.}
\end{video}  

{\bf Hydrodynamic model~-~} In order to understand these
time-dependent solutions and locate the regions of parameter space,
where they can occur, we now formulate a hydrodynamic model, determine
its stationary states and analyse their stability.  Our starting point
is the \xmg{momentum conservation equation} in the form
$\partial_t v_x=\partial_y\sigma_{xy}$.  For simplicity we only
consider a one-dimensional model, allowing for a velocity $v_x$ in the
flow direction, dependent on $y$ only.  In addition we introduce a
variable $w(y,t)$ for the internal state or the microstructure of the
fluid.  Such a variable has been introduced for many complex fluids,
such as liquid crystals, entangled polymer solutions and colloidal
systems~\cite{Olmsted2008}. Here we associate it with the frictional
contact network. In the simplest model we only consider a scalar
variable, representing e.g. the number of frictional
contacts~\cite{Mari2015,cates2013}, but are aware that a tensorial
quantity, such as the fabric tensor, might be more appropriate.  We
assume that the microstructure variable relaxes to a stationary state
$\partial_t w=(w-w^*)/\tau$ which should vanish in the absence of
stress $w^*(\sigma \to 0) \to 0$.  Furthermore dynamic rearrangements occur only due to driving,
so that $\tau^{-1}\propto \dot{\gamma}$. So far the model is the same
as considered by Nakanishi et al.~\cite{Nakanishi2012} for dilatant
fluids. However, the coupling of stress and microstructure is
different for the frictional grains under consideration. The
frictional contacts reduce the flow and hence the velocity
gradient. Starting from the constitutive equation for the strain rate
of frictionless grains $\dot{\gamma}_0= a\sigma^{1/2} + c\sigma^2$, we
take the strain rate of our system of particles with friction to be
$\dot{\gamma}=\dot{\gamma}_0-w$. This completes the definition of the
hydrodynamic model
\begin{eqnarray}
\label{dyn_model}
\partial_t \dot{\gamma}&=& \partial_y^2 \sigma\nonumber\\
\dot{\gamma}&=&\dot{\gamma}_0-w\nonumber\\
\partial_t w&=&-\frac{\dot{\gamma}}{\Gamma}(w-w^*). 
\end{eqnarray}
Here we have introduced a proportionality constant,
$\tau=\Gamma/{\dot{\gamma}}$, which can be fitted, when comparing
simulations to the predictions of the model. For the stability
analysis, it is irrelevant. Higher-order diffusive terms may be added
to the stress and/or the $w$-equation. We have checked that the
inclusion of such terms does not change \xmg{the stability analysis. Other
hydrodynamic models of granular fluids include velocity fluctuations 
\cite{Brilliantov},
e.g. granular temperature which, however cannot explain
effects due to friction.}

The model allows for two stationary states: The first one corresponds
to stationary flow and is explicitly given by
\begin{equation}
\dot{\gamma}=\dot{\gamma}_0-w^*; \quad w=w^*; \quad \sigma=\sigma_0
\end{equation}
Given that $w^*$ should vanish for vanishing shear, we take it as
$w^*=b\sigma$, so that we recover the constitutive relation for
frictional grains, Eq.~(\ref{eq:fc1}).  The second stationary solution
accounts for the jammed state and reads
\begin{equation}
\dot{\gamma}=0; \quad w=\dot{\gamma}_0; \quad \sigma=\sigma_0
\end{equation}
For both stationary states, the stress $\sigma=\sigma_0$ is
homogeneous over the sample, in agreement with the Navier-Stokes
equation which require a homogeneous stress in two dimensions and
hence do not allow vorticity banding.

\begin{figure}
  \includegraphics[angle=0, width=1.0\linewidth]{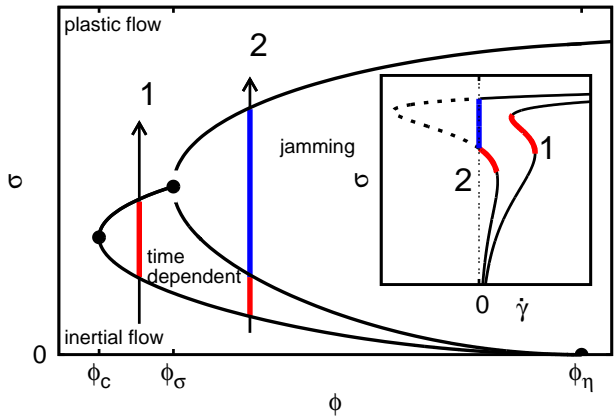}
  \caption{\label{fig:stability} Inset: Typical stationary states
    (flow curves) in the range $\phi_{\sigma}>\phi>\phi_c$ (1) and
    $\phi>\phi_{\sigma}$ (2). Unstable regions are highlighted in red
    (time dependent flow) and blue (jamming). Main: Phase
    diagram following from the linear stability analysis; the two
    generic flow curves, displayed in the inset,
    correspond to the paths, denoted by 1 and 2.}
\end{figure}

\begin{figure}
  \includegraphics[width=1.0\linewidth]{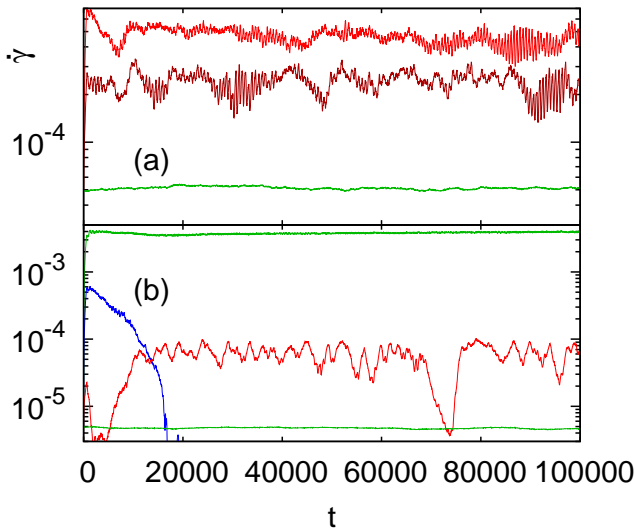}
  \caption{\label{fig:chaotic_strain}Strain rate $\dot{\gamma}(t)$ as a
    function of time for several values of stress, corresponding to
    path 1 (top, $\phi=0.7975$) and path 2 (bottom, $\phi=0.8035$) in Fig.~\ref{fig:stability}; the
    sudden drops of the strain rate in the time dependent flow curve
    indicate that the system nearly gets jammed.}
\end{figure}

{\bf Stability analysis~-~} In order to study the stability of the
stationary states, we consider small deviations
$\delta\sigma, \delta w \sim e^{\Omega t}e^{iky}$ and linearise
Eq.\ref{dyn_model} in $\delta\sigma, \delta w$. As expected the
stationary flow is unstable for
$\frac{\partial \dot{\gamma}}{\partial\sigma} < 0$. Below $\phi_c$,
this does not occur and we find two stable modes: a hydrodynamic one,
$\Omega_1\propto -k^2$ corresponding to the conservation of momentum,
and a nonhydrodynamic one
$\Omega_2\propto- \frac{\partial \dot{\gamma}}{\partial\sigma} \to 0$,
corresponding to the relaxation of the microstructure, whose
relaxation time becomes infinite as $\phi\to\phi_c$ and $\sigma \to \sigma_c$. Above $\phi_c$,
the model predicts two stable stationary flow solutions, inertial flow
at small stress and plastic flow at large stress. In between a gap of
unstable stress values occurs, such that no stationary homogeneous
flow is possible in this range of stresses. A typical flow curve in
the range \xmg{$\phi_c < \phi < \phi_{\sigma}$} is shown in the inset of
Fig.~\ref{fig:stability} as line 1, indicating the unstable regime as
red. The jammed state is only stable for $\phi>\phi_{\sigma}$ in the
region where the constitutive relation yields a negative
$\dot{\gamma}$. A typical flow curve in the range
\xmg{$\phi_{\sigma} < \phi < \phi_{\eta}$} is line 2 in the inset of
Fig.~\ref{fig:stability} with the stable jammed state marked in
blue. For $\phi>\phi_{\eta}$, only stationary plastic flow and
stationary jamming are predicted by linear stability analysis. The
resulting phase diagram of the model is shown in the main panel of
Fig.~\ref{fig:stability}.

\begin{figure}
  \includegraphics[angle=0, width=1.0\linewidth]{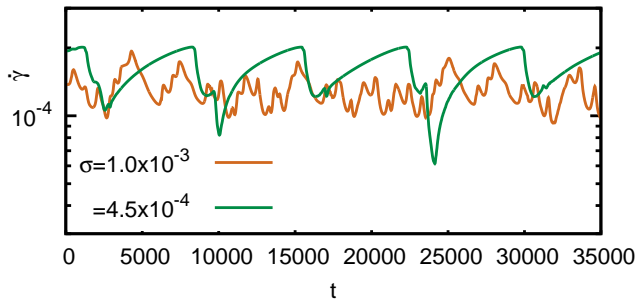}
  \caption{\label{fig:timedependence_model} Spatially averaged strain rate vs. time from
    numerically integrating the hydrodynamical model for different
    imposed stress. In the unstable region at intermediate
    stress-values we observe oscillations and chaotic
    solutions ($\Gamma=10^{-3}, \phi = 0.7975$).}
\end{figure}  

Above the critical packing fraction, $\phi_c$, a finite range of
inaccessible stress values with
$\frac{\partial\dot{\gamma}}{\partial\sigma}\leq 0$ exists and gives
rise to a corresponding range of unstable wavenumbers $k^2\leq
k_c^2=\left|\dot{\gamma}\frac{\partial
    \dot{\gamma}}{\partial\sigma}\right|$, which shrinks to 0 as
$\phi\to\phi_c$. Hence the model predicts an approximately
harmonically oscillating state at the onset of instability, while well
inside the unstable region more and more wavenumbers are unstable so
that one expects a broad range of frequencies to be present in the
spectrum. These expectations are born out by numerical integration of
the partial differential equations in order to obtain the full
nonlinear dynamical evolution. Close to $\phi_c$, the oscillations are
approximately harmonic, while we find oscillating and seemingly
chaotic solutions at larger packing fractions (see Fig.~\ref{fig:timedependence_model}).
 
{\bf Comparison with simulations~-~} To check the predictions of the
above analysis, we performed stress controlled simulations (for
technical details see Ref.~\cite{Grob2014}) along the paths 1 and 2 in
the phase diagram.  The time-dependent strain rate along path 1 is
shown in Fig.~\ref{fig:chaotic_strain}a for three different stress
values. \xmg{The lowest one shows stationary flow in the Bagnold
regime, the chaotic time-dependence with
oscillatory components is represented by the two red curves for intermediate stress} and larger stress gives rise to stationary
plastic flow (not shown in Fig.~\ref{fig:chaotic_strain}\xmg{a}). Similarly in
Fig.~\ref{fig:chaotic_strain}b we show the strain rate as a function of
time for path 2. In addition to the steady-state flow, and the
oscillatory flow, there are stationary jammed states, where the
initial flow ceases after a certain amount of time.

\begin{figure}
  \includegraphics[angle=0,width=0.9\linewidth]{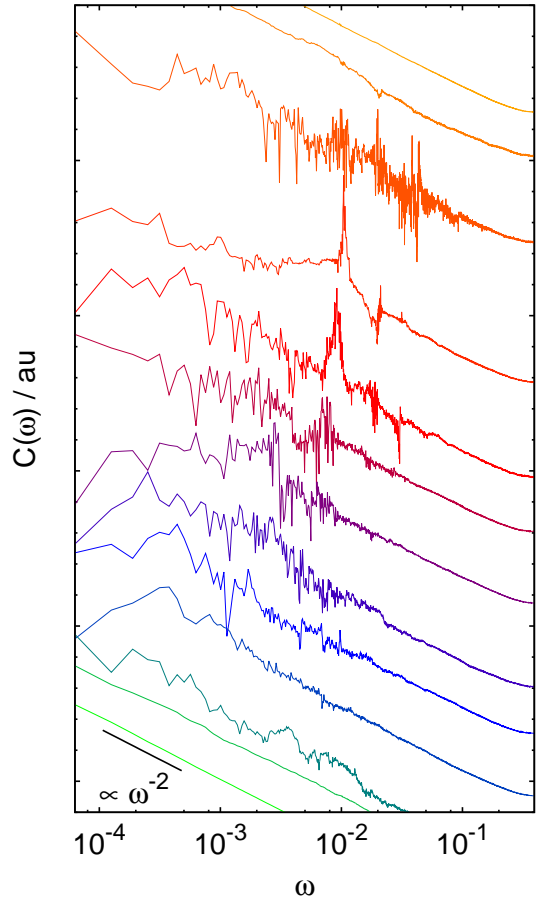}
  \caption{\label{fig:spectrum} Power spectrum $C(\omega)$ of strain rate
    fluctuations at constant stress from low (bottom) to high (top)
    values of $\sigma$ (shifted vertically for clarity of presentation); in the
    stable region $C\sim \omega^{-2}$, whereas in the unstable region
    additional complex structures are superimposed on the
    $\omega^{-2}$ decay.}
\end{figure}

To get a quantitative measure of the irregularity in the time
dependence of the states, we have computed the power spectrum
$C(\omega)$ as the Fourier transform of the strain rate
auto-correlation function (Fig.~\ref{fig:spectrum}). In the stable regimes, the power
spectrum is $C\sim \omega^{-2}$, suggesting simple exponential
correlations and linear noise. In the unstable regime this background
spectrum is superposed by additional and irregular complex
structures. This is a strong indication of truly nonlinear chaotic
dynamics. Noteworthy is also the strong peak at higher stresses. This
corresponds to the fast oscillations visible in
Fig.~\ref{fig:chaotic_strain}a. 

{\bf Finite-size effects~-~} Snapshots from video~\ref{fig:chaotic_local_stress} indicate the build-up of
large-scale coherent structures. A large correlation length has also
been observed in Ref.~\cite{heussinger2013} in the context of
continuous shear thickening in non-Brownian particle
suspensions. These correlations are also in line with the strong
finite-size effects observed in the flow curves of
Fig.~\ref{fig:flowcurveS}.  Indeed, lowering the system size the
oscillatory state acquires a finite lifetime and the system jams. This
is because in small systems there is a finite probability that large
strain rate fluctuations towards $\dot\gamma\to0$ lead into a stable
jammed state (see Fig.~\ref{fig:chaotic_strain}b for an example of
such an excursion). Similarly, the appearance of s-shaped flow curves
is due to system-size dependent strong strain rate-fluctuations towards
the jammed state with $\dot\gamma=0$.

{\bf Conclusion~-~}
We discuss stress-controlled driving of a granular system that
undergoes discontinuous shear thickening. In particular a regime is identified where the system does not settle \xmg{into a
time-independent steady-state.} Instead, it displays spatio-temporal
oscillations and chaotic behavior as a novel possibility to adopt to stress in the unstable parts of the flow curve.
\xmg{Recent experiments on corn-starch reveal very similar unsteady flow where theory predicts discontinuous shear thickening~\cite{hermes2015}.}

Simulations reveal a phase diagram that has a characteristic
re-entrant form with steady-state flow at small and large stresses.
At intermediate values of stress either time-dependent states are
observed or the system settles into a non-flowing jammed state,
depending on stress and packing fraction. For
$\phi_c<\phi<\phi_{\sigma}$ only time dependent solutions are
observed, whereas for $\phi_{\sigma}<\phi$ a sequence of inertial
flow, chaotic flow, jammed state and plastic flow are seen for
increasing values of $\sigma$, until at $\phi=\phi_{\eta}$
only a transition from the jammed state to plastic flow remains.

We also present a hydrodynamical model, coupling stress to a
microstructural observable. Within linear stability analysis we
recover the detailed features of the phase diagram as obtained from
simulations. In the unstable region, the model predicts either
oscillating or time-chaotic flow.

%
In future work we plan to quantify the spatio-temporal correlations
that are visible in the snapshots and compare them with length-scales
determined in \cite{heussinger2013} from velocity correlations. 
We furthermore aim to better understand the microstructural observable,
check whether it can be associated with the contacts which are blocked
by Coulomb friction and explore the possibility of a tensorial
observable, such as the fabric tensor.

\section{Acknowledgments}\label{acknowledgments}
We gratefully acknowledge financial support by the DFG via FOR 1394 and the Emmy Noether program (He 6322/1-1).

\appendix

\section{Appendix}

In this \xold{appendix} we deal with flow curves of frictional grains at fixed packing fraction when the particles' elastic and viscous damping constants change. 

In the main article we use unit strength for elastic and viscous contribution in the linear-spring dashpots ($k$ and $\eta$, respectively). Units of time and energy dissipation when particles interact are set by this choice. In particular, the coefficient of (normal) restitution, $\epsilon^{(n)}$, and the binary collision time, $t$, are set. Table \ref{tab:eps} summarizes the normal coefficient of restitution and the binary collision time for the parameters that we discuss here. We tune both contributions, $k=k^{(n)}=k^{(t)}$ and $\eta=\eta^{(n)}=\eta^{(t)}$, independently.
\begin{table}[h]
  \centering
  \caption{\label{tab:eps}Normal restitution coefficient $\epsilon^{(n)}$ and binary collision time $t$ for different $k$ and $\eta$.}
  \begin{ruledtabular}
    \begin{tabular}{cccc} 
      $k$ & $\eta$ & $\epsilon^{(n)}$ & $t$ \\
      \hline
      1 & 1 & 0.305 & 2.375 \\
      1 & 1/2 & 0.569 & 2.26 \\
      1 & 1/10 & 0.895 & 2.22 \\
      1/2 & 1 & 0.163 & 3.63 \\
      2 & 1 & 0.44 & 1.4 \\
      10 & 1 & 0.7 & 0.7 \\
    \end{tabular}
  \end{ruledtabular}
\end{table}
Both, $\epsilon^{(n)}$ and $t$, just serve as a guidance on properties of pairwise collisions and do not respect the large packing fraction. Also the important frictional contribution which implies tangential restitution $\epsilon^{(t)}$ is not characterized by these quantities. The normal restitution can be computed easily while the tangential part is of rich and complicated behaviour due to its dependence on the impact velocities \cite{becker2008}.

\begin{figure}[h]
  \centering
  \includegraphics[width=1.0\linewidth]{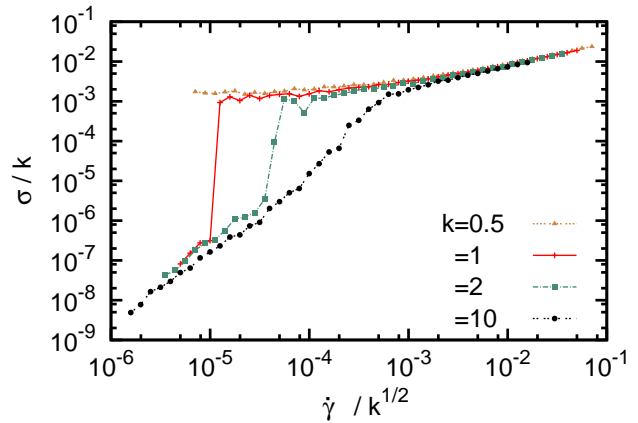}
  \caption{\label{fig:k}Scaled flow curves for $\phi=0.80$, viscous damping parameter $\eta=1$ and varying elastic constant $k$.}
\end{figure}

\begin{figure}[h]
  \centering
  \includegraphics[width=1.0\linewidth]{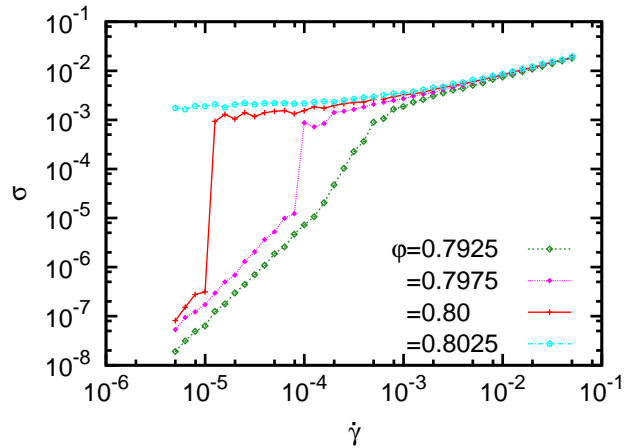}
  \caption{\label{fig:phi}Flow curves for varying packing fraction $\phi$ across the transition with viscous damping parameter $\eta=1$ and elastic constant $k=1$.}
\end{figure}

\begin{figure}[h]
  \centering
  \includegraphics[width=1.0\linewidth]{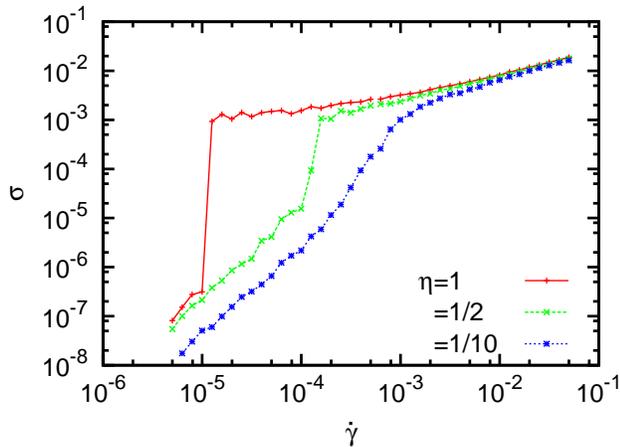}
  \caption{\label{fig:eta}Flow curves for $\phi=0.80$, elastic constant $k=1$ and varying viscous damping parameter $\eta$.} 
\end{figure}

The following data shows flow curves of a system with $N=8000$ particles. Fig.~\ref{fig:k} shows scaled flow curves for fixed $\phi=0.80$ and $\eta=1$. 
When $k$ decreases ($\epsilon^{(n)}$ decreases), the discontinuity shifts towards small strain rate until it is missed out by our simulation. Note that the numerical effort for low strain rate is much larger as for large strain rate. An increasing $k$ (increasing $\epsilon^{(n)}$) shifts the discontinuity towards larger strain rate which makes the discontinuity smaller until it vanishes. At $k=10$ the transition from inertial to plastic flow is smooth and without shear thickening. The phenomenology of varying $k$ seems in that range similar to a change of the packing fraction. The latter was studied in previous work \cite{Grob2014}. A systematic study of variations of all three parameters, $\phi, k$ and $\eta$, is out the scope of this article. In Fig.~\ref{fig:phi} we show a choice of packing fractions with phenomenology similar to Fig.~\ref{fig:k}.
Lower $\phi$ leads to flow curves similar to those with large $\epsilon^{(n)}$ and large $\phi$ show the same behaviour as small $\epsilon^{(n)}$.
The variation of the viscous damping parameter $\eta$ is shown in Fig.~\ref{fig:eta}.
The phenomenology is the same as above: decreasing $\epsilon^{(n)}$ (larger $\eta$) leads to a shift towards small strain rates and an increasing coefficient of restitution (smaller $\eta$) shifts the discontinuity towards larger strain rate until it vanishes. Then the transition between inertial and plastic flow is smooth and without shear thickening again.


\end{document}